\pdfoutput=1
\def\year{2017}\relax
\documentclass[letterpaper]{article} %DO NOT CHANGE THIS
\usepackage{aaai17}  %Required
\usepackage{times}  %Required
\usepackage{helvet}  %Required
\usepackage{courier}  %Required
\usepackage{url}  %Required
\usepackage{graphicx}  %Required

\usepackage{booktabs} 
\usepackage{subcaption}
\usepackage{amsfonts}  
\usepackage{mathtools}
\DeclarePairedDelimiterX{\infdivx}[2]{(}{)}{%
  #1\;\delimsize\|\;#2%
}

\usepackage{color,soul}
\newcommand{\vect}[1]{\boldsymbol{#1}}
\usepackage[linesnumbered, ruled]{algorithm2e}
\newcommand*{\affaddr}[1]{#1} % No op here. Customize it for different styles.
\newcommand*{\affmark}[1][*]{\textsuperscript{#1}}
\newcommand*{\email}[1]{\texttt{#1}}

\begin{document}
\title{Modeling Game Avatar Synergy and Opposition through Embedding in Multiplayer Online Battle Arena Games}
\author{%
Zhengxing Chen\affmark[1], Yuyu Xu\affmark[1],  Truong-Huy D. Nguyen\affmark[2], Yizhou Sun\affmark[3] \and Magy Seif El-Nasr\affmark[1]\\
\affaddr{\affmark[1]College of Information and Computer Science, Northeastern University}\\
\affaddr{\affmark[2]Department of Computer and Information Science, Fordham University}\\
\affaddr{\affmark[3]Department of Computer Science, University of California, Log Angeles}\\
\affaddr{\affmark[1] \email{czxttkl@gmail.com, yuyuxu@ccs.neu.edu, m.seifel-nasr@neu.edu}} \\
\affaddr{\affmark[2] \email{tnguyen88@fordham.edu}} \\
\affaddr{\affmark[3] \email{yzsun@cs.ucla.edu}} 
}
\maketitle

%In fact, the learning curve is especially steep for novice players who are still struggling with the game's control mechanics. 

% automatic team formation. lack of correspondance to later case study
%\huy{this means players could select a game avatar that they love to play, then there is a match-making system that puts them in a team, automatically}.  
%Moreover, team-based video games is a perfect testbed for team formation analysis, a crucial field in crowdsourcing, human resource management and robotics.
% Therefore, effective numerical encoding of game avatars' relationships that can aid players is highly desirable, which leads to a wide range of helpful downstream applications, including match outcome prediction and avatar pick recommender.
\begin{abstract}
Multiplayer Online Battle Arena (MOBA) games have received increasing worldwide popularity recently. In such games, players compete in teams against each other by controlling selected game avatars, each of which is designed with different strengths and weaknesses. Intuitively, putting together game avatars that complement each other (\textit{synergy}) and suppress those of opponents (\textit{opposition}) would result in a stronger team. In-depth understanding of synergy and opposition relationships among game avatars benefits player in making decisions in game avatar drafting and gaining better prediction of match events. However, due to intricate design and complex interactions between game avatars, thorough understanding of their relationships is not a trivial task. 

In this paper, we propose a latent variable model, namely \textit{Game Avatar Embedding} (GAE), to learn avatars' numerical representations which encode synergy and opposition relationships between pairs of avatars. The merits of our model are twofold: (1) the captured synergy and opposition relationships are sensible to experienced human players' perception; (2) the learned numerical representations of game avatars allow many important downstream tasks, such as similar avatar search, match outcome prediction, and avatar pick recommender. To our best knowledge, no previous model is able to simultaneously support both features. Our quantitative and qualitative evaluations on real match data from three commercial MOBA games illustrate the benefits of our model. 

%Further, to illustrate the utility of the model, we present a case study of a game avatar recommendation system which can recommend game avatars that enhance the current line-up while matching its owner's personal preferences.
\end{abstract}

\section{Introduction}

%\begin{figure}
%\centering
%\includegraphics[width=0.45\textwidth]{DOTA-2-Heros}
%\caption{Game avatar gallery in DOTA 2. MOBA games have rich choices of game avatars that players can select to play. There are sophisticated team synergy and counter effects between differently skilled game avatars.}
%\end{figure}

% Video games is a huge industry that attracts hundreds of milions of players each year. Reportedly, \textit{e-Sports}, a family of competitive video games, is estimated to have 188 million viewership and 748 million dollar worth market in 2015 and the numbers are expected to grow continuously~\cite{superdata}. Among them, 

%\chen{ the original writing is not accurate. it is concurrent at peak in the year 2014.}

Multiplayer Online Battle Arena (MOBA) has been one of the most popular e-sports game types. For example, \textit{League of Legends} (Riot Games), one of the most popular e-Sports, reportedly has 90 million accounts registered, 27 million unique daily players, and 7.5 million concurrent users at peak~\cite{fanbase,27million}. In a MOBA game, 2 teams, composed of 5 player each, combat in a virtual environment. The goal is to beat the opposite team and destroy their  base. Game avatars are often designed with a variety of attributes, skills, roles, etc., which is intended to provide players with choices and options so that every player can find a character that fits their preferences. Moreover, it is customary for avatars in such games to possess strength in one aspect, but weakness in others. As such, in order to win a match, it is well-known that players need to not only control their own game avatars well, but also need to select a game avatar that, together with other team members' picks, forms a team that enhances skills and complements weaknesses of each other (\textit{synergy}), while posing suppressing strengths over those in the opponent team (\textit{opposition})\footnote{\url{http://www.weskimo.com/a-guide-to-drafting.html}} \footnote{\url{https://www.reddit.com/r/learndota2/comments/3f9szo/how_to_counter_pick_heroes/}} 
. For example, in DOTA 2 (Valve Corporation), avatar \textit{Clockwerk} has high synergy with \textit{Naix} because Clockwerk can transport Naix to target enemy directly, increasing the limited mobility of Naix to hit enemy. Naix also delivers large damage to complement the limited attack by Clockwerk, making them an efficient fighting duo. In another example, avatar \textit{Anti-Mage}'s mana burn skill reduces an opponent's mana resource, making him a natural opposition to \textit{Medusa} whose durable capability is completely relying on how much mana it has.

% game avatars with the \textit{Agility} attribute such as \textit{Anti-Mage} excel in long-range physical damage but are susceptible in face-to-face combats. In contrast, with sturdier armor, those of the \textit{Strength} attribute such as \textit{Axe} can tolerate large damage in direct confrontation but lack strong attack ability}.  
 
Comprehensive understanding of synergy and opposition relationships between game avatar enhances player awareness and experience in games. First, it allows players to make good decisions in drafting their team's game avatars to maximize the chance of winning. Second, it improves the prediction of the match's progress and final outcomes, which helps players in preparing for strategies in advance. Lastly, it helps players discover other game avatars that match their personal expertise or preferences. However, due to the intricate design and complex interactions among game avatars, thorough understanding of game avatars' pros and cons and their relationships is not a trivial task to human players.

%also important as it will guide players to select game avatars and take strategic actions. 
%\huy{the paragraph below appears forced. I comment it out for now.}
%\st{Therefore, explainable, interpretable analysis of game avatar characteristics is needed because it can aid players understand the game in depth. Such model should answer many questions which "black-box" algorithms fail to answer...} 
%For example, \textit{Given avatar A works well with avatar B, which avatar else works well with B as well?} \textit{Which avatars are most similar to avatar C?} (See Section ~\ref{mq} for more motivational questions.) Apparently, answering these questions can spark many novel application directions in the video game industry, such as game avatar clustering, and intelligent recommendation systems for selecting game avatars.

In order to model game avatars' synergy and opposition relationships, we propose a latent variable model, called \textit{Game Avatar Embedding} (GAE). GAE models game avatars as vectors in a low-dimensional space learned. We hypothesized that the probability function of a match outcome constitutes of pairwise synergy and opposition interactions formulated by game avatar vectors. Game avatar vectors and other model parameters are learned by gradient descent through maximizing the likelihood function of all observed match outcomes. Latent variable models (LVM) and embedding techniques have been shown to successfully capture characteristics of entities in texts \cite{mikolov2013distributed}, graphs \cite{maaten2008visualizing}, and recommendation systems \cite{koren2009matrix}. The advantages of these techniques include: 1. less manual feature engineering required, 2. robust learning even when the data sparsity problem is present, and 3. better reusability for downstream tasks than pure predictive algorithms such as Gradient Boosting Decision Tree~\cite{friedman2001greedy} and Factorization Machine~\cite{rendle2010factorization}.

Inheriting the advantages from LVM, GAE is the first model to our best knowledge that not only captures synergy and opposition relationships robustly, but also facilitates many important downstream tasks which consume game avatar vectors as input, e.g., similarity search on game avatars, team composition analysis~\cite{kim2016proficiency,agarwala2014learning} and match outcome prediction~\cite{yang2016real}.

%
%In this paper, we choose to use data from a typical family of team-based multiplayer video games, Multiplayer Online Battle Arena (MOBA), which has been increasingly popular worldwide in recent years~\cite{fanbase,27million} \huy{you could have introduced MOBAs earlier in the paper; team-based games can mean a lot of vastly different game play, such as sports games like Fifa so it could be a bit too vague}.  MOBAs are typically characterized by rich choices of game avatars and dynamic strategies for team cooperation and counter. Therefore, it is an ideal test bed to study how game avatar characteristics and synergism and antagonism between them affect a match's outcome. Nevertheless, we expect that our model can also be applied directly or with minimal modification to other types of team-based multiplayer video games, such as team-based \textit{First Person Shooting} (FPS) and sport games.

% Although we focus on video game data in this paper, it is worth noting that understanding team synergy and counters is also valuable for team formation in other fields such as social networks~\cite{lappas2009finding,anagnostopoulos2012online}, robotics~\cite{liemhetcharat2012modeling}, crowdsourcing~\cite{rahman2015worker,kittur2010crowdsourcing,roy2015task}, education~\cite{buckenmyer2000using} and human resource management~\cite{Mohrman}. Our model can therefore be applicable to discovering team member characteristics and their synergy in the aforementioned domains.

In sum, the contributions of our paper are multi-fold:
\begin{enumerate}
\item we describe a novel Game Avatar Embedding (GAE) model which characterizes game avatars in vectors in terms of synergy and opposition;
\item we demonstrate the effectiveness of the model via quantitative experiments on real data from three commercial MOBA games;
\item we showcase how our model facilitates downstream tasks such as similar avatar search and avatar pick recommendation, which off-the-shelf Machine Learning models cannot accomplish. 
\end{enumerate}

Our paper is structured as follows. We first present related work, then introduce GAE models with details on model specification and learning process. Next, we report quantitative evaluation results of GAE, followed by two case studies. Finally, we conclude our paper and discuss limitations as well as future directions.

\section{Related Works}
As we adopt an embedding model to uncover synergy and opposition relationships among game avatars from team-based outcomes for MOBA games, our work aligns with recent research in team formation analysis, MOBA game research, and embedding-based modeling methods.

\subsection{Team Formation Analysis}
Team formation analysis is the research topic which aims to uncover factors that impact team performance. Team formation has already been studied in MOBA games \cite{pobie1,Semenov2016}. \cite{pobie1} verified that team success depends on a successful selected combination of game avatars.  \cite{Semenov2016} predicts MOBA outcomes by using 2-way factorization machine \cite{rendle2010factorization}, which can reliably estimate the levels of pairwise relationships through factorized parameterization. However, their method does not naturally derive meaningful numerical representations of game avatars that can be analyzed and utilized for many other downstream applications.

Although we focus on video game data in this paper, team formation analysis could help advance many other domains, such as social networks~\cite{lappas2009finding,anagnostopoulos2012online}, crowdsourcing~\cite{rahman2015worker,kittur2010crowdsourcing,roy2015task},  and robotics \cite{liemhetcharat2012modeling}. Existing works that also use machine learning to learn chacacterization of team members  \cite{liemhetcharat2012modeling,rahman2015worker} are different than our work in that: (1) the dimensions of team member characterization are often pre-defined and fixed, such as a fixed set of skills, which requires manual efforts and domain knowledge; (2) no opposition relationship has been modeled.   
 
\subsection{MOBA Game Research}
The rich design of MOBA games has attracted variety of research to be conducted upon them. For example, team formation analysis~\cite{pobie1,pobie2,neidhardt2015team,kim2016proficiency,agarwala2014learning}, skill decomposition~\cite{zhengxing2016player}, match outcome prediction and avatar pick recommendation systems~\cite{bhattacharyadata}. They shed lights on real-world problems or facilitate building adaptive player experience~\cite{chen2015analytics}. Many of these tasks rely on processing vectors which encode characteristics of game avatars. For example, in team formation analysis, the calculation of team diversity is averaged pairwise cosine distances between game avatars' attribute vectors. Principal Component Analysis~\cite{jolliffe2002principal} and t-SNE~\cite{maaten2008visualizing}, two frequently used dimension reduction techniques in clustering and visualization, are also based on entities' vectors. Our GAE model induces the vectors of game avatars encoding their synergy and opposition relationships, which can facilitate many downstream tasks that perform upon vectors.

\subsection{LVMs and Embedding Models}
LVMs/embedding models have been long studied in Natural Language Processing (NLP) \cite{mikolov2013distributed}, graph \cite{maaten2008visualizing}, and recommendation system \cite{koren2009matrix}. In this family of models, entities are associated with vectors in a shared, continuous low-dimensional space which encode entities' characteristics efficiently and effectively. We will use vectors and embedding interchangeably to refer to the numerical representations of entities.

% Entity vectors can be learned in either supervised~\cite{collobert2011natural}, unsupervised~\cite{mikolov2013distributed,mikolov2013efficient}. For supervised learning, the values of entity vectors are learned towards specific prediction tasks. This results in highly task-related characteristics being extracted. Unsupervised learning allows entity representation learned from co-occurrence information from massive unlabeled data and often results in more general representation.

Some salient advantages of embedding models/LVMs are as follows: 1. it requires little human labor for feature engineering because entity vectors can be learned based on labels or what are observed explicitly (e.g., links between nodes in a graph, user-item matrix, word sequences). 
2. information between entities can be shared more effectively during the learning phase. For example, similar embeddings of two similar words can be learned if they are often used and occur in the similar contexts, even though they do not appear together. 3. learned vector representation can be reused by many kinds of applications, such as sentiment analysis \cite{maas2011learning} and data visualization \cite{maaten2008visualizing}. 

In our paper, game avatars are embedded as low-dimensional vectors. Their values are learned (supervised) through maximizing the winning probabilities (defined in Section 3) of all observed match outcomes. We will show in \textit{Performance Evaluation} and \textit{Case Study} that the learned game avatar embeddings indeed capture sensible team-related characteristics and allow for other downstream applications, such as similar avatar search and avatar pick recommendation. This cannot be achieved by previous methods such as Logistic Regression, Factorization Machine~\cite{Semenov2016} and Gradient Boosting Decision Tree~\cite{friedman2001greedy} which simply predict match outcomes without a means to derive game avatar embeddings to be reused in other tasks. 

% without a means to derive deeper insights on how the composition of avatars from both teams affects the outcomes. Our game avatar embedding sheds some light on this aspect by revealing how synergistic and opposition effects can help explain the outcomes.

%\hl{bilinear model}
%In practice, there can be built many kinds of application based on learned vectors of entities. 
%In latent factor model, vectors are usually learnt in supervised way by approximating some objective function. in embedding models, unsupervised are more common. Use contexts information to obtain vectors.
%NLP embedding can be seen in PTE paper.

\section{Preliminary and Problem Definition}

% \subsection{Problem Definition}
Suppose the training data is a match set $\mathcal{M}=\{M_1, M_2, \cdots, M_Z\}$ with $Z$ matches. There are $N$ unique game avatars appearing in total, denoted by $\mathcal{A}=\{A_1, A_2, \cdots, A_N\}$. We assume each match is competed between two teams, the red and the blue team. We use $\mathcal{T}_{z, r}=\{A_{i}\}$ and $\mathcal{T}_{z, b}=\{A_{j}\}$ to denote the sets of game avatars in the red team and the blue team in $M_z$, respectively. Since we are studying 5-vs-5 MOBA games, we have for $\forall z$, $|\mathcal{T}_{z, r}|=5$ and $|\mathcal{T}_{z, b}|=5$. We use $\mathcal{T} = \{(\mathcal{T}_{z,r}, \mathcal{T}_{z,b}) | z = 1, \cdots, Z \}$ to denote all game avatar line-ups of $\mathcal{M}$.

Match outcomes are marked as $\vect{O}=\{o_1, o_2, \cdots, o_Z\}$. $o_z=1$ means the red team wins over the blue team in $M_z$ otherwise $o_z=0$.  We use $p(o_z=1)$ and $p(o_z=0)$ to denote the winning probability from the view of the red team and the blue team, respectively. Hence, $p(o_z=0)=1-p(o_z=1)$. 

\section{Game Avatar Embedding Model}
In this section, we will describe the proposed model, the learning process, as well as discuss its relationships with Factorization Machines, a related model.

\subsection{Model Synergy and Opposition}

Inspired by embedding methods which have managed to learn low-dimensional vectors to capture abundant attributes of entities, we propose to map characteristics of game avatars into a low-dimensional latent space. For a game avatar $A_i$, its latent feature vector is denoted as $\vect{a}_i \in \mathbb{R}^K$. $\vect{A} \in \mathbb{R}^{N \times K}$ is the latent feature matrix such that $\vect{A}=\{\vect{a}_i\}$. 

We choose to use a bilinear model to model synergy and opposition relationships between pairs of avatars. The bilinear model allows us to separately learn game avatar embeddings, as well as the matrices that determine the extents of synergy and opposition across different dimensions of game avatar embeddings. 

First, we introduce the intra-team \textit{synergy score function} $S(i, j)$, which calculates the level of synergy to which $A_i$ exerts on $A_j$ in the same team:
\begin{equation}
\begin{aligned}
S(i, j) = \vect{a}_i^T \cdot \vect{P} \cdot \vect{a}_j  = \sum\limits_{m=1}^K\sum\limits_{n=1}^K a_{im} \cdot p_{mn} \cdot a_{jn}
\label{eqn:syn}
\end{aligned}
\end{equation}

$\vect{P} \in \mathbb{R}^{K \times K}$ is named \textit{synergy matrix}. There are two ways to understand $\vect{P}$ intuitively:
\begin{enumerate}
\item one can think of $\vect{a}_i^T \cdot \vect{P} = \vect{a}_i'$ as converting $A_i$'s embedding into the dimensions that $A_j$ looks for as a helpful teammate. Then, the higher the dot product is between $\vect{a}_i'$ and $\vect{a}_j$, the higher synergy the two game avatars can build. 
\item alternatively, one can think that $p_{mn}$ measures how much $m$-th dimension of $\vect{a}_i$ fits $n$-th dimension of $\vect{a}_j$ in terms of intra-team interaction. 
\end{enumerate}

Second, we define the inter-team \textit{opposition score function} $C(i, j)$, which quantifies the extent to which $A_i$ counters $A_j$ in the opposite team:
\begin{equation}
\begin{aligned}
C(i, j) = \vect{a}_i^T \cdot \vect{Q} \cdot \vect{a}_j = \sum\limits_{m=1}^K\sum\limits_{n=1}^K a_{im} \cdot q_{mn} \cdot a_{jn}
\label{eqn:ant}
\end{aligned}
\end{equation}

$\vect{Q} \in \mathbb{R}^{K \times K}$ is named \textit{opposition matrix}. In a similar way to understand $\vect{P}$, $q_{mn}$ measures the influence on $A_i$ countering $A_j$, given their embeddings' interaction on $m$-th and $n$-th dimension respectively.

%Moreover, for better interpretability of the model, we add the constraints that for $\forall i$, $\vect{a}_i \geq 0$ . The constraints eliminate the undesirable situations that can be illustrated in the following example. Suppose $p_{mn}>0$, which can be interpreted that $m$-th dimension and $n$-th dimension interact positively for synergism. Also, suppose that $a_{im} < 0$ and $a_{jn}<0$, which means $\vect{a}_i$ and $\vect{a}_j$ are relatively weak in $m$-th and $n$-th dimension respectively. Intuitively, the pair of the dimensions should \textit{barely} contribute to $S(i, j)$, contrary to the fact that $a_{im} \cdot p_{mn} \cdot a_{jn} > 0$.

Note that $\vect{P}$ and $\vect{Q}$ are not necessarily symmetric, as the level of opposition in which $A_i$ suppress $A_j$ could be different from that of $A_j$ on $A_i$. 

In this model, we only capture pairwise relationships because they are much more prevalent. We also find advanced models such as Gradient Boosting Decision Trees~\cite{friedman2001greedy} potentially considering more intricate relationships do not improve the match outcome prediction task on all the three datasets we study (See Section~\textit{Performance Evaluation}). Still, it is possible to extend GAE for higher order interactions by modeling them using tensors and tensor operations~\cite{kolda2009tensor}. We will explore this aspect in the future.

\subsection{Model Winning Probability}

Next, we propose to model \textit{a winning outcome as the linear breakdown of the individual biases, as well as their intra-team and inter-team interactions, of game avatars from the two teams involved.} Individual biases represent game avatars' intrinsic control difficulty that affects match outcomes, denoted as $\vect{b}=\{b_1, b_2, \cdots, b_N\}$. Hence, the winning outcome $p(o_z=1)$ of a match $M_z$ is defined as follows: 
 
\begin{equation}
\begin{aligned}
p(o_z=1) = \sigma \big( & \sum\limits_{i \in \mathcal{T}_{z,r}} b_{i} - \sum\limits_{j \in \mathcal{T}_{z,b}} b_{j} \\
& + \sum\limits_{i, j \in \mathcal{T}_{z,r} \atop{i \neq j}} S(i, j) -  \sum\limits_{i, j \in \mathcal{T}_{z,b} \atop{i \neq j}} S(i, j)\\
& + \sum\limits_{i \in \mathcal{T}_{z,r}}\sum\limits_{j \in \mathcal{T}_{z,b}} C(i, j) - C(j, i) \big)
\end{aligned}
\label{eqn:wp}
\end{equation}
where $\sigma(\cdot)$ is sigmoid function $\frac{1}{1+exp(-x)}$.

The input of the sigmoid function is the sum of the differences of: (1) individual biases towards winning, (2) synergy strength inside the team, and (3) opposition intensity against the opponent team. The latter two differences depend on traversing all valid pairs of game avatars within the same team or across the two teams. The larger the differences are, the closer $p(o_z=1)$ is to 1 meaning the advantageous team is more likely to win.

Note that in our formulation, players' individual skill levels are not accounted for in the winning outcome's probability. This is reasonable, since the data we collected is from highly selective ranked matches. Most commercial MOBA games have proprietary matchmaking systems to ensure that only sufficiently experienced players with similar skill levels are allowed to compete in ranked matches (the type of matches we study). Therefore, the chance of results being skewed by data from incompetent players is low. % for example dota2 only allows level 20+ players to play ranked games.

\subsection{Objective Function and Learning} 
Assuming that each match is independent, the overall likelihood function is:
\begin{equation}
p(\mathcal{O}, \mathcal{T}|\vect{A}, \vect{P}, \vect{Q}, \vect{b}) = \prod\limits_{z=1}^{Z} p(o_z=1)^{o_z}p(o_z=0)^{1-o_z}
\end{equation}

The objective function is to minimize the negative log likelihood w.r.t $\vect{\Theta} = \{\vect{A}, \vect{P}, \vect{Q}, \vect{b}\}$:
\begin{equation}\label{eqn:constobj}
\begin{aligned}
J(\vect{\Theta}) & =-\frac{1}{Z} \sum\limits_{z=1}^Z \big(o_z \log p(o_z=1) + (1-o_z) \log p(o_z=0) \big) 
\end{aligned}
\end{equation}

For parameter learning, we use AdaGrad \cite{duchi2011adaptive} to update parameters based on a small batch of matches in each iteration. 
  
\subsection{Relation to Factorization Machine Model}
% We want to state that, \textit{while FM and GAE can both be formed as a bilinear model, there is no obvious way for FM to extract game avatar embedding from its model parameters.}

GAE has a close relationship with 2-way factorization machine (FM) \cite{rendle2010factorization}, which has been applied in \cite{Semenov2016} to predict match outcomes of the same kind of games. In \cite{Semenov2016}, for a match $M_z$, the feature vector $\vect{x}_z \in \{0,1\}^{2N}$ is a binary vector indicating which five avatars appear in the red and blue team respectively:

 \begin{equation}
    x_{zi}=
    \begin{cases}
      1, & \text{if } i \leq N \text{ and avatar } i \text{ was in the red team} \\
      & \text{or if } i > N \text{ and avatar } i-N \text{ was in the blue team} \\
      0, & \text{otherwise}
    \end{cases}
   \label{eqn:featcons}
  \end{equation}

and FM models a winning probability by additionally exploring pairwise interactions between non-zero features: 
\begin{equation}
\begin{aligned}
p(o_z=1) =& \sigma \big(\sum\limits_{i \in \mathcal{T}_{z,r}} c_{i} + \sum\limits_{j \in \mathcal{T}_{z,b}} c_{j+N} \\
& + \sum\limits_{i, j \in \mathcal{T}_{z,r} \atop{i < j}} <\vect{v}_i, \vect{v}_j> + \sum\limits_{i, j \in \mathcal{T}_{z,b} \atop{i < j}} <\vect{v}_{i+N}, \vect{v}_{j+N}>\\
& + \sum\limits_{i \in \mathcal{T}_{z,r}}\sum\limits_{j \in \mathcal{T}_{z,b}} <\vect{v}_i, \vect{v}_{j+N}> \big),
\end{aligned}
\label{eqn:wpfm}
\end{equation}
{\setlength{\parindent}{0cm}
where $c_i \in \mathbb{R}$ and $\vect{v}_i \in \mathbb{R}^K$ for $\forall i=1,\cdots ,2N$ are first-order and second-order parameters, and $<\cdot, \cdot>$ is dot product operation. Therefore, each avatar $A_i$ is associated with a quartet of parameters $(c_i, c_{i+N}, \vect{v}_i, \vect{v}_{i+N})$.
}

%We can show that by using Eqn.~\ref{eqn:rewritesyn} and \ref{eqn:rewritecnt}, the winning probability function of GAE  can be formatted into that of the factorization machine. This can be achieved by 

Dot productions in Eqn.~\ref{eqn:wpfm} can be re-written in the vector-matrix-vector product form that is similar to Eqn.~\ref{eqn:syn} and Eqn.~\ref{eqn:ant}. For each $A_i$, we can set $\vect{v}_i = U_i \vect{u}_i$ and $\vect{v}_{i+N} = V_i \vect{u}_{i}$ where $U_i$ and $V_i$ are two  matrices that linearly transform the \textit{same} base $\vect{u}_i$. $\vect{u}_i$ can be seen as the equivalence of $\vect{a}_i$ in GAE, a vector capturing characteristics of $A_i$. Therefore, each pairwise term in Eqn.~\ref{eqn:wpfm} can be written as:

%for unique pair $i, j$ from the red team:
 \begin{equation}
    \begin{aligned}
    <\vect{v}_i, \vect{v}_j> &= \vect{u}_i^T  (U_i^T U_j) \vect{u}_j \\
    \end{aligned}
   \label{eqn:fmijred}
  \end{equation} 

%for unique pair $i, j$ from the red team:  
  \begin{equation}
    \begin{aligned}
    <\vect{v}_{i+N}, \vect{v}_{j+N}> &= \vect{u}_i^T  (V_i^T V_j) \vect{u}_j \\
    \end{aligned}
   \label{eqn:fmijblue}
  \end{equation} 

%for unique pair $i, j$ from opposite teams:  
  \begin{equation}
    \begin{aligned}
    <\vect{v}_{i}, \vect{v}_{j+N}> &= \vect{u}_i^T  (U_i^T V_j) \vect{u}_j \\
    \end{aligned}
   \label{eqn:fmijopposite}
  \end{equation}

FM and GAE both attempt to estimate second-order interactions through the embedding/factorization technique. Therefore, they both inherit the advantages of the factorization technique that all pairs of co-occurrences can help the learning of any particular pair of interaction. Since the hierarchy of both models is the linear summation of first-order biases and second-order interactions, FM and GAE are expected to have similar classification performance in match outcome prediction. However, for FM, it is unclear how to determine $U_i$ and $V_i$ for $\forall i$. On the contrast, GAE by design can learn game avatar embedding and synergy and opposition matrices at the same time, which enables practitioners to reuse game avatar embedding for other downstream tasks.
% The explicit modeling of game avatar embeddings by GAE also allows the objective function to incorporate more aspects of information in the future. For example, one can add unsupervised learning based objective \cite{tang2015pte}, human perceived constraints \cite{wagstaff2001constrained}, or side information \cite{singh2008relational} to improve the quality of embedding learning.

\section{Performance Evaluation}\label{ressection}

% \subsection{Data Sets}
We evaluate the utility of GAE using datasets collected from three commercial MOBA games, namely Defense of the Ancients (DOTA2), Heroes of Newerth (HoN), and Heroes of the Storms (HotS). All data is from 5-vs-5 matches that pit ten random players in two teams against each other. No major game update affecting the mechanics of the games occurred during the data collection phase. All three games employ matchmaking systems that select only players of similar skill levels when assembling a match. All the three datasets have roughly balanced winning outcomes for both the red and blue teams. Statistics of the three datasets is shown in Table \ref{tab:freq}. 

The HotS match data was downloaded from a third-party game log website\footnote{\url{https://www.hotslogs.com/Info/API}}; all the matches happened during the last month of 2016. The HoN dataset was collected by~\cite{suznjevic2015application}, which contains matches played between December 20, 2014 from April 29, 2015. Finally, for DOTA 2, we use  the original data set collected between February 11, 2016 to March 2, 2016 by Semenov et al.~\cite{Semenov2016}, and extract a subset of matches played by gamers with similar skill levels (i.e., normal level). 
% Dictated by the nature of their abilities and conceived preferences, game avatars are categorized into Carry, Nuker, Initiator, Disabler, Durable, Escape, Jungler, Support and Pusher.

\begin{table}
  \caption{Statistics of datasets}
  \label{tab:freq}
  \begin{tabular}{lccc}
    \toprule
    & HotS & HoN & DOTA2\\
    \midrule
    \# of Matches & 1,814,066 & 1,101,046 & 3,056,596 \\
    \# of Avatars & 58 & 126 & 111 \\
  \bottomrule
\end{tabular}
\end{table}

% dota role categorization
% http://dota2.gamepedia.com/Role
% http://wiki.teamliquid.net/dota2/Hero_Roles

% All used data and codes are open sourced and available online \footnote{\url{https://github.com/czxttkl/GAE}}.

\subsection{Experiment Setup and Results}

There are two experiments designed to assess the effectiveness of GAE. The first is conducted as a numerical evaluation in terms of outcome prediction, while the second evaluates GAE's interpretability using human experts, as compared to other state-of-the-art methods.

\subsection{Outcome Prediction Results}

First, we evaluate the match outcome prediction performance of GAE against well-known baselines, including Logistic Regression (\textbf{LR})\footnote{\url{http://scikit-learn.org/}}, Gradient Boosting  Decision Trees (\textbf{GBDT})\footnote{\url{https://github.com/dmlc/xgboost}} and 2-way Factorization Machine (\textbf{FM})\footnote{\url{https://github.com/ibayer/fastFM}}.  
For each game dataset, we adopt 10-fold cross-validation procedure with train:validate:test ratio set to be 8:1:1. In each fold, a model with different configurations of hyperparameters (e.g.,  regularization penalty, the number of trees, the dimension of latent space, etc.) is trained on the train dataset and the best hyperparameters is determined according to the classification performance on the validation dataset. The classification performance of the model with the best hyperparameters on the test dataset will be recorded as the final measurement of its classification strength. For \textbf{GAE}, we use Eqn. ~\ref{eqn:wp} to predict outcomes on test datasets. 
For baseline models, Eqn.~\ref{eqn:featcons} is used to construct feature vectors, similar to how it is done in previous works. 
The area under ROC Curve (AUC) is used as the classification performance measurement. Ten test AUC are recorded during the 10-fold cross-validation for each model (LR, GBDT, FM and GAE) such that classification performance can be compared using paired t-test (with confidence level 0.001).

\begin{table}
  \caption{Outcome prediction AUC on test datasets; \textbf{(*)} indicate where GAE outperforms with $p$-values $< 0.001$.}
  \label{tab:auc}
  \begin{tabular}{c@{\hskip 0.5in}c@{\hskip 0.36in}c@{\hskip 0.36in}c@{\hskip 0.36in}c}
    \toprule
    Models  & HotS & HoN  & DOTA2  \\
    \midrule
    LR   & 0.6095\textbf{*}    & 0.6115\textbf{*}    & 0.6875\textbf{*}     \\
    GBDT & 0.6375\textbf{*}    & 0.6144\textbf{*}    & 0.7014\textbf{*}   \\
	FM   & 0.6440    & 0.6154\textbf{*}    & 0.7143   \\
    \midrule
    GAE  & 0.6437    &  0.6220    &  0.7143 \\
    %GAE-NB &         &           & 0.7146    &   0.6602 \\
  \bottomrule
\end{tabular}
\end{table}

Table~\ref{tab:auc} reports the classification performances of all models in match outcome prediction. The paired t-tests showed that GAE has significantly higher test AUC than other models except GAE vs. FM in Hots and DOTA2. 

We observe that LR has the worst classification AUC in all three games. That is not surprising because LR does not model interactions between avatars. This verifies that there do exist team synergy and opposition between game avatars. GBDT is a tree-based model that could handle interactions among more than two game avatars. However, it achieves statistically worse results than GAE. This demonstrates: (1) the strength of embedding methods in effectively encoding meaningful information of pairwise synergy and opposition relationships in a low-dimensional space; (2) much more data might be needed for GBDT to fully capture more complicated relationships. When GAE and FM are tuned with a proper number of latent space dimensions $K$, they achieve comparable AUC in HotS and DOTA2. This verified our expectation in Section \textit{Relation to Factorization Machine Model} that GAE and FM should have similar outcome prediction performance because they both rely on factorization techniques to quantify pairwise interactions. However, the exception is HoN where GAE is statistically significantly better than FM in HoN and GAE appears to have smaller improvement over LR than in other games. We will investigate the characteristics of HoN compared to other MOBA games in the future. Overall, GAE predicted match outcomes well and robustly.

\begin{table}
  \caption{Pearson's $r$ between human ratings and GAE/baseline scores on 3 evaluation sets of pairs (boldface indicates $p$-values $< 0.001$)}
  \label{tab:pearson}
  \begin{tabular}{cccc}
    \toprule
    & Similarity & Synergy & Opposition \\
    \midrule
    Win-Ratio Matrix & 0.5258 & - & - \\
    FM & - & \textbf{0.7841} & \textbf{0.7669} \\
    \midrule
    GAE & \textbf{0.8080} & \textbf{0.8488} & \textbf{0.7384}  \\
  \bottomrule
\end{tabular}
\end{table}

\subsection{Human Evaluation}\label{hl}
%One notable strength of GAE is its ability to learn reasonable game avatar embeddings, which could not be obtained by other models.

Second, we would like to validate how sensible GAE results are as compared to the experts', i.e., human players', judgment. We ask human players to rate pairs of game avatars in terms of similarity, synergy and opposition. Intuitively, if a model's scores are highly correlated with human ratings, we conclude that such model generates sensible results. Since recruitment of knowledgeable players is a relatively expensive task, we only evaluate on the DOTA2 dataset. Based on a pilot test of three DOTA2 players, 60 pairs are selected which have clear similarity, synergy and opposition relationships (20 pairs for each kind of relationship). For example, the 20 similarity evaluation pairs include both very similar as well as very different pairs of game avatars because either kind is expected to be evaluated consistently by subjects. 

When using GAE to evaluate the pairs, the similarity is determined by cosine similarity between the learned game avatar embeddings. The synergy is determined by $S(i, j) + S(j, i)$ and the opposition by the absolute value of $C(i, j) - C(j, i)$ for any pair of game avatars $A_i$ and $A_j$. Note that besides GAE, we are not aware of any approach that can handle similarity, synergy, and opposition queries all in a single model. FMs can naturally answer synergy and opposition queries; more specifically, two avatars' synergy and opposition levels can be obtained using the left hand side of Eqn.~\ref{eqn:fmijred} and Eqn.~\ref{eqn:fmijopposite} respectively. 
However, they are not designed for similarity search, so we created an ad-hoc baseline method to compute avatars' similarity based on the cosine similarity between the respective rows of a win-ratio matrix $W \in \mathbb{R}^{N \times 2N}$, constructed as:
\begin{equation}
W_{i,j} = \frac{\text{\# of matches } (A_i, A_j) \text{ win}}{\text{\# of matches } (A_i, A_j)  \text{ from the same team}}
\end{equation}
\begin{equation}
W_{i, j+N} = \frac{\text{\# of matches } A_i \text{ wins over } A_j}{\text{\# of matches } A_i, A_j \text{ from the opposite teams}}
\end{equation}

To collect human ratings, we created a survey asking subjects to rate on a 5-point Likert scale the level of similarity, synergy or opposition on the 60 pairs, with 1 as ``not at all" and 5 as ``very much", and asked ten similarly skillful DOTA2 players to provide their ratings. 
%The survey was distributed to two related game forums \footnote{\url{https://www.reddit.com/r/DotA2/} and \url{https://steamcommunity.com/app/570/discussions/}}. 
%The pairs in the survey were shuffled each time to eliminate the potential bias due to the order of reading. 
%We also collected skill information from players along the survey. At last, each rating is the average of 10 different players who fall into the same skill range as our training data. 
We produce Pearson's $r$ between human ratings and GAE/baseline scores on the 20 pairs in each kind of relationship.

We compared the correlations (using Pearson's $r$) between human ratings and those by GAE and the baseline. Better correlation corresponds to more sensible results from the players' perspective.  
As shown in Table~\ref{tab:pearson},
for similarity queries, GAE's results better correlate with human ratings than those by the baseline, suggesting that similarity search based on the learned embeddings by GAE are more sensible. For synergy and opposition queries, both GAE and the baseline correlate with human ratings with high Pearson's $r$ ($\geq 0.7$) with $p$-value $<0.001$, which indicates both methods are sensible to human players. This can be explained by the similarity of FM and GAE's approach in using the factorization/embedding technique to model pairwise interactions.

%\begin{table}[]
%\centering
%\caption{Similarity, synergy and counter search results by GAE, with respective values}
%\label{tab:casestudy}
%\scalebox{0.9}{
%\begin{tabular}{|l|l|}
%\hline
%\textbf{Queried Avatar} & \textbf{Similarity} ($\text{cos\_sim}(\vect{a}_i, \vect{a}_j)$) \\ \hline
%Clinkz & Weaver (0.41), Riki (0.33), Mirana (0.30)                 \\ 
%Anti-Mage       & Juggernaut (0.33), Faceless Void (0.30), \\
% & Alchemist (0.25) \\ \hline
%\vhtable
%\textbf{Queried Avatar} & \textbf{Synergy} ($\vect{a}_i^T \vect{P}_{s} \vect{a}_j$) \\ \hline
%Vengeful Spirit & Drow Ranger (17.96), Luna (7.44), \\
%& Medusa (7.44)            \\ 
%Axe             & Omniknight (10.64), Ancient   \\
%&  Apparition (8.56), Dazzle (8.02)       \\ \hline
%\vhtable
%\textbf{Queried Avatar} & \textbf{Opposition} ($\vect{a}_i^T \vect{Q}_{s} \vect{a}_j$) \\ \hline
%Earthshaker     & Meepo (23.51), Phantom Lancer (19.63),  \\
%& Chaos Knight (18.93)  \\ 
%Phantom Lancer  & Wraith King (28.23), Viper (18.46) \\
% & Huskar (14.99)           \\ \hline
%\end{tabular}
%}
%\end{table}

\section{Case Study}\label{cs}
In this section, we would like to qualitatively demonstrate the quality and utility of GAE within the context of practical applications. All analyses are done with the help of three seasoned DOTA2 players, conducted on GAE's results ($K=75$) as learned from DOTA2 data. Game avatars are called \textit{heroes} in DOTA 2.

% Before proceeding, we describe some characteristics of DOTA2 avatars (\textit{heroes}) to facilitate the understanding of the content below. According to design attributes, heroes can be divided into three types: Strength, Agility and Intelligence. According to the demand for in-game resources, heroes are categorized as Support, Carry, Mid-, and Off-laner. According to roles in team combats and strategies, heroes are divided into Nuker, Disabler, Durable, Escaper, etc. As these categorizations are from different perspectives, they can overlap with each other. For instance, many Carry heroes have high Agility attribute, most Support are Intelligence heroes, and most Strength heroes play a role of Durable in combat. All these perspectives together contribute to the play style of a hero.

\subsection{Application - Similarity Search}
One direct downstream application utilizing GAE's game avatar vectors is similarity search. It can help players, both starters or pros, to expand their hero pools by recommending heroes similar to what they are already familiar with or good at. For example, given input hero \textit{Clinkz}, the top three heroes GAE returns are \textit{Weaver}, \textit{Riki} and \textit{Mirana}. After examining the results, the three seasoned DOTA2 players all agreed that the top three heroes are very similar to Clinkz, as they are all \textit{Agility Carry} heroes with low hit points, great escape capability, and sharing a stealthy play style.

\subsection{Application - Personalized Recommendation}
Kim et al.~\cite{kim2016proficiency} suggested that the ideal game avatar to maximize the winning chance fit players' personal expertise and team congruency in parallel, guided by which GAE could be used for a personalized avatar pick recommendation system. We select a real match played by one of our DOTA2 players for illustration although the real implementation and verification of this idea requires more work in the future. 

In a ranked match, the player is the last to pick a hero, when his team have picked \textit{Puck}, \textit{Ember Spirit}, \textit{Lion}, \textit{Necrophos}, and the opposite team picked \textit{Silencer}, \textit{Pudge}, \textit{Sand King}, \textit{Juggernaut}, \textit{Anti-Mage}. Given 30 seconds to make the pick, he wants to prioritize the hero selection that synergizes with his team and opposes the other team. Using Eqn.~\ref{eqn:wp} to search the hero that maximizes the winning probability, GAE returns the top recommendation,  \textit{Ursa}. However, the player has not played Ursa before, thus are less confident about playing it. Based on the similarity search on the learned game avatar embeddings, GAE returns a list of heroes similar to Ursa, top 3 being \textit{Troll Warlord}, \textit{Sven} and \textit{Juggernaut}. Finally, the player decides to go for Sven since that is the hero he is experienced with and Sven is also one of the top 5 heroes identified with best overall synergy and opposition besides Ursa.

Analyzing the above example, all the three DOTA2 players strongly agreed that Ursa is a very suitable choice given that this player's team lacks burst physical damage. In addition, they had different extra interpretations on the Ursa pick. For example, one player identified that Ursa could help the player finish the game early, disallowing the opposing team to elongate the game when Anti-mage will show his max advantage as a \textit{late game Carry}. Another player recognized that Ursa could increase team fights capability since Ursa is a \textit{Tank Carry} hero who is durable in fights. All the three players agreed that Sven is similar to Ursa, as both heroes output high burst physical damage. They also proposed that GAE recommendation can be used differently according to personal prioritization. For example, some players who strongly prefer skill familiarity can first use GAE to list their familiar heroes then run synergy and opposition search.
 
As a summary, GAE provides an interface to perform queries of similar, synergy and opposition simultaneously. These capabilities can then be incorporated into downstream applications, giving users a white-box tool to help them better understand the game and make better in-game choices that maximize the winning chance. 

\section{Conclusions, Limitations and Future Works}
Modeling synergy and opposition relationships between game avatars is an important task that helps players understand the game and make better decisions in forming effective teams. To tackle this task, our proposed embedding-based method models synergy and opposition relationships with game avatars encoded as vector representation. Our quantitative and qualitative analyses show that GAE is able to capture pairwise synergy and opposition relationships between game avatars that are sensible to human players. Moreover, the learned game avatar embeddings effectively capture important characteristics of game avatars because similarity search based on the game avatar embeddings also highly correlate with human ratings. Our model opens new doors to many downstream tasks, such as similarity search on game avatars and personalized avatar recommendation.

There are some future directions that we will pursue next. First, we want to study the extension of GAE in capturing higher-order relationships that involve more than two avatars, as well as study the trade-offs between its performance improvement and computation overhead incurred. Second, we are currently limited to access to sensitive human player information. In the future, we hope to collect match data with richer player information and model player and game avatar embeddings within the same model.

\bibliographystyle{aaai}
\bibliography{sigproc} 

\end{document}